\documentclass[12pt]{article}
\usepackage{amsmath}
\usepackage{bm}
\usepackage{amssymb}
\usepackage{amsthm}
\usepackage{longtable}
\usepackage{amsfonts}
\usepackage{geometry}
\usepackage{multirow}
\usepackage{multicol}
\usepackage{graphicx}
\usepackage{epsfig}
\usepackage{epstopdf}
\usepackage{subfigure}
\usepackage{color}
\usepackage{ulem}
\usepackage{booktabs}
\usepackage{rotating}
\usepackage[toc,page]{appendix}
\usepackage{cite}
\usepackage{cases}
\usepackage{makecell}
\allowdisplaybreaks[4]

\renewcommand{\theequation}{\arabic{section}.\arabic{equation}}

\begin{document}



\def\a{\frac{\alpha}{\eta}}
\def\b{\beta}
\def\d{\delta}
\def\e{\epsilon}
\def\g{\gamma}
\def\h{\mathfrak{h}}
\def\k{\kappa}
\def\l{\lambda}
\def\o{\omega}
\def\p{\wp}
\def\r{\rho}
\def\t{\tau}
\def\s{\sigma}
\def\z{\varsigma}
\def\x{\xi}
\def\V={{{\bf\rm{V}}}}
 \def\A{{\cal{A}}}
 \def\B{{\cal{B}}}
 \def\C{{\cal{C}}}
 \def\D{{\cal{D}}}
\def\G{\Gamma}
\def\K{{\cal{K}}}
\def\O{\Omega}
\def\R{\bar{R}}
\def\T{{\cal{T}}}
\def\L{\Lambda}
\def\f{E_{\tau,\eta}(sl_2)}
\def\E{E_{\tau,\eta}(sl_n)}
\def\Zb{\mathbb{Z}}
\def\Cb{\mathbb{C}}

\def\R{\overline{R}}

\def\beq{\begin{equation}}
\def\eeq{\end{equation}}
\def\bea{\begin{eqnarray}}
\def\eea{\end{eqnarray}}
\def\ba{\begin{array}}
\def\ea{\end{array}}
\def\no{\nonumber}
\def\le{\langle}
\def\re{\rangle}
\def\lt{\left}
\def\rt{\right}

\baselineskip=20pt

\newfont{\elevenmib}{cmmib10 scaled\magstep1}
\newcommand{\preprint}{
   \begin{flushleft}
   \end{flushleft}\vspace{-1.3cm}
   \begin{flushright}\normalsize
   \end{flushright}}
\newcommand{\Title}[1]{{\baselineskip=26pt
   \begin{center} \Large \bf #1 \\ \ \\ \end{center}}}

\newcommand{\Author}{\begin{center}
	\large \bf
	Pengcheng Lu${\,}^{a}$,
    Junpeng Cao${\,}^{b,c,d,e}$,
    Wen-Li Yang${\,}^{e,f,g}$,
    Ian Marquette${\,}^{h}$
    and Yao-Zhong Zhang${\,}^{a}$
\end{center}}

\newcommand{\Address}{\begin{center}

    ${}^a$ School of Mathematics and Physics, The University of Queensland, Brisbane, QLD 4072, Australia\\
    ${}^b$ Beijing National Laboratory for Condensed Matter Physics, Institute of Physics, Chinese Academy of Sciences, Beijing 100190, China\\
	${}^c$ School of Physical Sciences, University of Chinese Academy of Sciences, Beijing 100049, China\\
	${}^d$ Songshan Lake Materials Laboratory, Dongguan, Guangdong 523808, China\\
	${}^e$ Peng Huanwu Center for Fundamental Theory, Xian 710127, China\\
    ${}^f$ Institute of Modern Physics, Northwest University, Xian 710127, China\\
    ${}^g$ Shaanxi Key Laboratory for Theoretical Physics Frontiers, Xian 710127, China\\
    ${}^h$ Department of Mathematical and Physical Sciences, La Trobe University, Bendigo, VIC 3552, Australia

\end{center}}

\Title{Exact surface energies and boundary excitations of the Izergin--Korepin model with generic boundary fields }
\Author

\Address
\vspace{1cm}

\begin{abstract}
\noindent The Izergin--Korepin model is an integrable model with the simplest twisted quantum affine algebra $U_q(A_2^{(2)})$ symmetry. Applying the $t$-$W$ method, we derive the homogeneous zeroes Bethe ansatz equations and the corresponding zeroes patterns of the Izergin--Korepin model with generic integrable boundaries. Based on these results, we analytically compute the surface energies and boundary excitations in different regimes of boundary parameters of the model. It is shown that in some regimes, correlation effect appears between two boundary fields.

\vspace{1truecm} \noindent {\it PACS:} 75.10.Pq, 03.65.Vf, 71.10.Pm

\end{abstract}
\newpage
\section{Introduction}
\label{intro} \setcounter{equation}{0}
The Izergin--Korepin (IK) model\cite{CMP791981} is a quantum integrable model \cite{PRL191967,Baxter1982} with the twisted quantum affine algebra $U_q(A^{(2)}_2)$ as underlying algebraic structure. It was shown that this model is closely related to the Dodd--Bullough--Mikhailov, also known as Jiber--Mikhailov--Shabat model \cite{PRSLA3521977,SPD241979}, which is a low-dimensional relativistic quantum field model. The IK model with periodic boundary conditions, domain-wall boundary conditions, or diagonal boundaries has been extensively studied \cite{NPB4351995,SPJETP571983,TMP761988,J9412192,NPB4501995,NPB3721992,NPB4881997,NPB5561999,NPB5581999,NPB5962001,LMP622002,NPB6702003,JSM2016}.
The open boundary case is related to the $O(n)$ loop model \cite{NPB4351995,IJMPB041990,JSM2013,JSM2017} and  self-avoiding walks  \cite{PRL741995,JPA462013,JSM2016,PRE952017}.
The IK model with generic integrable boundaries defined by non-diagonal reflection matrices  \cite{J9412192,NPB5581999,LMP622002} is not $U(1)$ symmetric and its exact solution was obtained in \cite{JHEP062014} by means of the so-called off-diagonal Bethe ansatz (ODBA).
However, the physical properties of the model remain an unsolved problem due to the non-homogeneity of the Bethe ansatz equations (BAEs) and the unclear patterns of the Bethe roots.

In \cite{PRB1022020,PRBL1032021}, a systematic $t$-$W$ $method$ for calculating physical properties of quantum integrable systems with or without $U(1)$-symmetry has been developed. In this method, the eigenvalues of the transfer matrix are characterized by zeroes, instead of the traditional Bethe roots. From the fusion relations, one can obtain the homogeneous Bethe ansatz equations for the zeroes. Taking the logarithm then the derivative of the Bethe ansatz equations, the exact thermodynamic properties and physical quantities of the systems in the thermodynamic limit can be calculated. Subsequently, this method was applied to many quantum integrable models, such as the
integrable $J_1$-$J_2$ model with competition interactions \cite{NPB9752022}
 and the $D_2^{(1)}$ spin chain with generic non-diagonal boundary reflections \cite{NPB9842022}. The algebraic structures of these models are related to untwisted affine algebras.
In this paper, we generalize the $t$-$W$ method to an integrable spin chain model with twisted affine algebra symmetry. We derive the density of zeroes and surface energy of the IK model with generic boundary fields using the $t$-$W$ method.

The paper is organized as follows.  Section 2 presents an introduction to the IK model and its integrability. In section 3, we derive the homogeneous zeroes BAEs by using the intrinsic properties of the transfer matrix and the zeroes of its eigenvalues. The BAEs completely determine the eigenvalues and energies of the system. In section 4, we give the detailed patterns of the zeroes in the different regimes of boundary parameters.
Section 5 is devoted to the computation of the surface energies induced by the generic boundary fields. The boundary excitations associated with the
boundary magnetic fields also are calculated in section 6. Concluding remarks are given in section 7. Some supporting materials are given in appendices A and B.


\section{Hamiltonian and  integrability}
\label{IK-H&I} \setcounter{equation}{0}

The Hamiltonian of the Izergin-Korepin (IK) model with generic integrable boundaries reads
\bea\label{IK-Ham}
H=H_{bulk}+H_{R}+H_{L}.
\eea
Here $H_{bulk}$ is the bulk Hamiltonian
\bea
H_{bulk}&\hspace{-0.2truecm}=\hspace{-0.2truecm}&\frac{2}{\sinh 5\eta-\sinh\eta} \sum_{j=1}^{N-1}\left\{
\cosh5\eta (E_{j}^{11}E_{j+1}^{11}+E_{j}^{33}E_{j+1}^{33})
\right. \nonumber \\[6pt]
&&\quad \left.+\sinh2\eta(\sinh3\eta-\cosh3\eta)
(E_{j}^{11}E_{j+1}^{22}+E_{j}^{22}E_{j+1}^{33})
\right. \nonumber \\[6pt]
&&\quad \left. +\sinh2\eta(\sinh3\eta+\cosh3\eta)
(E_{j}^{22}E_{j+1}^{11}+E_{j}^{33}E_{j+1}^{22})
\right. \nonumber \\[6pt]
&&\quad \left.+2\sinh\eta \sinh2\eta (e^{-2\eta}
E_{j}^{11}E_{j+1}^{33}+e^{2\eta} E_{j}^{33}E_{j+1}^{11})+\cosh\eta
(E_{j}^{13}E_{j+1}^{31}+ E_{j}^{31}E_{j+1}^{13})
\right. \nonumber \\[6pt]
&&\quad \left. +\cosh3\eta
(E_{j}^{12}E_{j+1}^{21}+E_{j}^{21}E_{j+1}^{12}
+E_{j}^{22}E_{j+1}^{22}+E_{j}^{23}E_{j+1}^{32}
+E_{j}^{32}E_{j+1}^{23}) \right. \nonumber \\[6pt]
&&\quad \left.  -e^{-2\eta}\sinh2\eta
(E_{j}^{12}E_{j+1}^{32}+E_{j}^{21}E_{j+1}^{23}) +
e^{2\eta}\sinh2\eta (E_{j}^{23}E_{j+1}^{21}+E_{j}^{32}E_{j+1}^{12})
\right\},\label{Ham}
\eea
where $\eta$ is the crossing parameter and $E_{j}^{\mu \nu}(\mu, \nu=1,2,3)$ is the Weyl matrix on $j$th-site with
$E^{\mu \nu}=|\mu \rangle \langle \nu |$, and
$H_L$ and $H_R$ quantify the left and right boundary fields, respectively,
\bea\label{LRboundary}
H_L=&&\hspace{-0.5truecm}2e^{-\varepsilon}[1+2e^{-\varepsilon}\sinh\eta]^{-1}\Big[\sinh\eta(E_1^{11}-E_1^{33} )+\cosh\eta E_1^{22}-e^{i\varsigma }E_1^{13}-e^{-i\varsigma }E_1^{31}  \Big],\no\\[6pt]
H_R=&&\hspace{-0.5truecm}2\Big[\sinh5\eta-\sinh\eta)(2e^{-\varepsilon^{\prime}}\sinh5\eta-4e^{-\varepsilon^{\prime}}\sinh\eta\cosh4\eta+1+2\cosh2\eta)\Big]^{-1}\no\\[6pt]
&&\hspace{-0.4truecm}\times\Big\{\Big[(e^{2\eta}-2e^{-4\eta-\varepsilon^{\prime}}\sinh\eta)\cosh5\eta+(1+2e^{-\varepsilon^{\prime}}\sinh5\eta)\sinh2\eta(\sinh3\eta-\cosh3\eta)\no\\[6pt]
&&\hspace{-0.4truecm}+2(e^{-4\eta}-2e^{2\eta-\varepsilon^{\prime}}\sinh\eta)\sinh\eta\sinh2\eta\Big]E_{N}^{11}+\Big[ (e^{2\eta}-2e^{-4\eta-\varepsilon^{\prime}}\sinh\eta)\sinh2\eta\no\\[6pt]
&&\hspace{-0.4truecm}\times(\sinh3\eta+\cosh3\eta)+(1+2e^{-\varepsilon^{\prime}}\sinh5\eta)\cosh3\eta+(e^{-2\eta}-2e^{4\eta-\varepsilon^{\prime}}\sinh\eta)\no\\[6pt]
&&\hspace{-0.4truecm}\times\sinh2\eta(\sinh3\eta-\cosh3\eta) \Big]E_{N}^{22}+\Big[2(e^{4\eta}-2e^{-2\eta-\varepsilon^{\prime}}\sinh\eta)\sinh\eta\sinh2\eta\no\\[6pt]
&&\hspace{-0.4truecm}+(1+2e^{-\varepsilon^{\prime}}\sinh5\eta)\sinh2\eta(\sinh3\eta+\cosh3\eta)+(e^{-2\eta}-2e^{4\eta-\varepsilon^{\prime}}\sinh\eta)\cosh5\eta\Big]E_{N}^{33}\no\\[6pt]
&&\hspace{-0.4truecm}-2e^{-\varepsilon^{\prime}}\sinh6\eta\cosh\eta(e^{2\eta+i\varsigma'}E_{N}^{13}+e^{-2\eta-i\varsigma'}E_{N}^{31}   )\Big\}\no\\[6pt]
&&\hspace{-0.4truecm}-\frac{2e^{-\varepsilon^{\prime}}(2\sinh\eta\sinh4\eta-\cosh5\eta)}{2\cosh2\eta-4e^{-\varepsilon^{\prime}}\sinh\eta\cosh4\eta+1+2e^{-\varepsilon^{\prime}}\sinh5\eta}I_N,
\eea
where $\varepsilon$, $\varepsilon^{\prime}$, $\varsigma$ and $\varsigma^{\prime}$ are the boundary parameters. We should note that $H_L$ and $H_R$ are two unparallel boundary fields, which obviously break the $U(1)$ symmetry. The hermiticity of the Hamiltonian (\ref{IK-Ham}) requires that the parameters $\eta$, $\varepsilon$, $\varepsilon^{\prime}$,  $\varsigma$ and $\varsigma^{\prime}-2i\eta$ are real. In this paper, we consider the case that the crossing parameter $\eta>0$.

Throughout, ${\rm \mathbf{V}}$ denotes a three-dimensional linear space, and let $\{|i\rangle | i=1,2,3\}$ be an orthonormal basis of it. We adopt standard notation: for any matrix $A \in {\rm End}({\rm \bf V})$, $A_j$ is an embedding operator in the tensor space ${\rm \bf V} \otimes {\rm \bf V} \otimes \cdots $, which acts as $A$ in the $j$-th space and as the identity on the other factor spaces; For $B \in {\rm End}({\rm \bf V} \otimes {\rm \bf V})$, $B_{ij}$ is an embedding operator of $B$ in the tensor space, which acts as the identity on all factor spaces except for the $i$-th and $j$-th ones.

The Hamiltonian (\ref{IK-Ham}) is constructed by using the $R$-matrix, and the type II  generic non-diagonal $K$-matrices $K^{L}(u)$ and $K^{R}(u)$ classified in \cite{NPB5581999,LMP622002}. The $R$-matrix $R(u)\in {\rm End}({\rm\bf V}\otimes {\rm\bf V})$ is $U_q(A_2^{(2)})$ symmetric and has the form \cite{CMP791981}
 \bea
 R_{12}(u)=\left(\begin{array}{r|r|r}{\begin{array}{rrr}h_3(u)&&\\&h_2(u)&\\&&h_4(u)\end{array}}
           &{\begin{array}{lll}&&\\e(u)&{\,\,\,\,\,\,\,\,\,\,\,\,\,\,\,\,\,\,\,}&{\,\,\,\,\,\,\,\,\,\,}\\{\,\,\,\,\,}&{\,\,\,\,}g(u)&{\,\,\,\,\,\,\,}\end{array}}
           &{\begin{array}{lll}&&\\&&\\f(u){\,\,\,\,\,\,\,\,\,}&{\,\,\,\,\,\,\,\,\,\,}&{\,\,\,\,\,\,\,\,\,\,}\end{array}}\\[12pt]
 \hline {\begin{array}{rrr}&\bar{e}(u){\,\,\,}&\\&&\bar{g}(u){\,\,}\\&&\end{array}}&
           {\begin{array}{ccc}h_2(u)&&\\&h_1(u)&\\&&h_2(u)\end{array}}
           &{\begin{array}{lll}&&\\g(u){\,\,}&&\\&e(u){\,\,\,\,\,\,\,\,\,\,\,\,\,\,\,\,}&\end{array}}\\[12pt]
 \hline {\begin{array}{ccc}&&\bar{f}(u){\,\,}\\&&\\&&\end{array}}
           &{\begin{array}{ccc}&{\,\,}\bar{g}(u){\,\,}&\\&&\bar{e}(u){\,\,}\\&&\end{array}}
           &{\begin{array}{ccc}h_4(u)&&\\&h_2(u)&\\&&h_3(u)\end{array}} \end{array}\right),\label{R-matrix}
\eea
where
\bea
&&h_1(u)=\sinh(u\hspace{-0.04truecm}-\hspace{-0.04truecm}3\eta)
\hspace{-0.04truecm}-\hspace{-0.04truecm}\sinh 5\eta\hspace{-0.04truecm}+\hspace{-0.04truecm}
\sinh 3\eta\hspace{-0.04truecm}+\hspace{-0.04truecm}\sinh\eta,\,\,
h_2(u)=\sinh(u\hspace{-0.04truecm}-\hspace{-0.04truecm}3\eta)
\hspace{-0.04truecm}+\hspace{-0.04truecm}\sinh3\eta,\no\\[6pt]
&&h_3(u)=\sinh(u-5\eta)+\sinh\eta,\quad h_4(u)=\sinh(u-\eta)+\sinh\eta,\no\\[6pt]
&&e(u)=-2e^{-\frac{u}{2}}\sinh2\eta\cosh(\frac{u}{2}-3\eta),\quad \bar{e}(u)=-2e^{\frac{u}{2}}\sinh2\eta\cosh(\frac{u}{2}-3\eta),\no\\[6pt]
&&f(u)=-2 e^{-u+2\eta}\sinh\eta\sinh2\eta-e^{-\eta}\sinh4\eta,\no\\[6pt]
&&\bar{f}(u)=2 e^{u-2\eta}\sinh\eta\sinh2\eta-e^{\eta}\sinh4\eta,\no\\[6pt]
&&g(u)=2e^{-\frac{u}{2}+2\eta}\sinh\frac{u}{2}\sinh 2\eta,\quad \bar{g}(u)=-2e^{\frac{u}{2}-2\eta}\sinh\frac{u}{2}\sinh 2\eta.
\label{R-element-2}
\eea
It satisfies the quantum Yang-Baxter equation (QYBE) \cite{PRL191967,Baxter1982}
\bea
R_{12}(u_1-u_2)R_{13}(u_1-u_3)R_{23}(u_2-u_3)
=R_{23}(u_2-u_3)R_{13}(u_1-u_3)R_{12}(u_1-u_2).\label{QYB}\eea
and the properties:
\bea
&&\hspace{-1.5cm}\mbox{ Initial
condition}:\hspace{0.17cm}\,R_{12}(0)= (\sinh\eta-\sinh5\eta)P_{12},\label{Int-R}\\
&&\hspace{-1.5cm}\mbox{ Unitarity
relation}:R_{12}(u)R_{21}(-u)= \varphi_1(u)\,\times {\rm id},\label{Unitarity}\\
&&\hspace{-1.5cm}\mbox{ Crossing
relation}:\,R_{12}(u)=V_1R_{12}^{t_2}(-u+6\eta+i\pi)V^{-1}_1,
\label{crosing}\\
&&\hspace{-1.5cm}\mbox{ PT-symmetry}:\hspace{0.57cm}\,R_{21}(u)=R^{t_1\,t_2}_{12}(u),\label{PT}\\
&&\hspace{-1.5cm}\mbox{ Periodicity}: \,
\qquad\hspace{0.27cm} R_{12}(u+2i\pi)=R_{12}(u),\label{Periodic}
\eea
where $P_{12}$ is the usual permutation operator and the function $\varphi_1(u)$ and the crossing matrix $V$ are given by
\bea
\varphi_1(u)=\hspace{-0.6truecm}&&-4\sinh(\frac{u}{2}-2\eta)\sinh(\frac{u}{2}+2\eta)\cosh(\frac{u}{2}-3\eta)\cosh(\frac{u}{2}+3\eta),\label{p-1-function}\\[6pt]
V=\hspace{-0.6truecm}&&\left(
          \begin{array}{ccc}
             &  & -e^{-\eta} \\
             & 1 &  \\
            -e^{\eta} &  &  \\
          \end{array}
        \right),\qquad V^2=1.
\eea

The type II generic non-diagonal $K$-matrices $K^{L}(u)$ and $K^{R}(u)$ are given by \cite{J9412192} \footnote{The $K$-matrices belong to one of the two types of general non-diagonal $K$-matrices obtained in \cite{NPB5581999,LMP622002}, namely, those which were classified as type II there. They are very generic $K$-matrices. By solving the reflection equation of the IK model, it is found that there are two types of solutions. The first kind of solution is that all the elements in the reflection matrix take the non-zero values. However, these matrix elements are not independent and there are only two free parameters. This solution is defined as the Type I reflection matrix. The second kind of solution is that the non-diagonal matrix elements $k_{12}(u)$, $k_{21}(u)$, $k_{23}(u)$ and $k_{32}(u)$ are zero. These kind of solution is classified as the Type II reflection matrix which still have two free parameters. These two types of reflection matrices are inequivalent and cannot be obtained from each other through any transformation.}
\bea
&&K^{L}(u)=\left(
          \begin{array}{ccc}
            1+2e^{-u-\varepsilon}\sinh\eta & 0 & 2e^{-\varepsilon+i\varsigma}\sinh u \\
            0 & 1-2e^{-\varepsilon}\sinh(u-\eta) & 0 \\
            2e^{-\varepsilon-i\varsigma}\sinh u & 0 & 1+2e^{u-\varepsilon}\sinh\eta \\
          \end{array}
        \right),\label{Kf}\\[6pt]
&&K^{R}(u)={\cal M}K^{L}(-u+6\eta+i\pi)\Big|_{(\varepsilon,\varsigma)\rightarrow(\varepsilon^{\prime},\varsigma^{\prime})},\label{Kz}
\eea
where ${\cal M}$ is the diagonal constant matrix ${\cal M}=diag(e^{2\eta},1,e^{-2\eta})$. $K^{L}(u)$ satisfies the reflection equation (RE)
\bea\label{RE}
R_{12}(u-\mu)K_1^{L}(u)R_{21}(u+\mu)K_2^{L}(\mu)=K_2^{L}(\mu)R_{12}(u+\mu)K_1^{L}(u)R_{21}(u-\mu),
\eea
and $K^{R}(u)$ satisfies the dual RE
\bea\label{DRE}
R_{12}\hspace{-0.7truecm}&&(\mu-u)K_1^{R}(u){\cal M}_1^{-1}R_{21}(-u-\mu+12\eta){\cal M}_1K_2^{R}(\mu)\no\\
\hspace{-0.7truecm}&&=K_2^{R}(\mu){\cal M}_2^{-1}R_{12}(-u-\mu+12\eta){\cal M}_2K_1^{R}(u)R_{21}(\mu-u).
\eea

We now define the single-row monodromy matrices
$T(u)$ and $\hat{T}(u)$, which are two $3\times 3$ matrices with operator-valued elements acting on the quantum space ${\rm\bf V}^{\otimes N}$,
\bea
T_0(u)=&&\hspace{-0.6truecm}R_{0N}(u-\theta_N)R_{0\,N-1}(u-\theta_{N-1})\cdots R_{01}(u-\theta_1),\label{Mon-V-1}\\
\hat{T}_0(u)=&&\hspace{-0.6truecm}R_{10}(u+\theta_1)R_{20}(u+\theta_{2})\cdots R_{N0}(u+\theta_N),\label{Mon-V-2}
\eea
where ${\rm\bf V}_0$ is a $3\times 3$ auxiliary space.
The corresponding transfer matrix $t(u)$ is given by
\bea
t(u)= tr_0\{K^R_0(u)T_0(u)K^L_0(u)\hat{T}_0(u)\},\label{trans}
\eea
The Hamiltonian (\ref{IK-Ham}) then can be expressed in terms of the transfer matrix $t(u)$ as
\bea
\hspace{-0.8truecm}H=&&\hspace{-0.6truecm}-\frac{\partial \,\ln t(u)}{\partial u}\Big|_{u=0,\{\theta_j=0\}}.\label{Ham-1}
\eea

The QYBE (\ref{QYB}), the RE (\ref{RE}) and its dual (\ref{DRE}) give rise to the commutation relation $[t(u),t(v)]=0$ \cite{JPA198821}, which ensures the integrability of the Hamiltonian (\ref{IK-Ham}).

\section{Homogeneous BAEs}
\label{Homogeneous-BAEs} \setcounter{equation}{0}

Using the properties of the $R$-matrix (\ref{R-matrix}) and the $K$-matrices (\ref{Kf})-(\ref{Kz}), the following operator identities can be easily proved
\bea
&&t(\pm \theta_j)t(\pm \theta_j+6\eta+i\pi)=\frac{\delta_1(u)\times {\rm id}}{\varphi_1(2u)}\Big|_{u=\pm\theta_j},\label{hds1}\\[6pt]
&&t(\pm \theta_j)t(\pm \theta_j+4\eta)=\frac{\delta_2(u)\times t(u+2\eta+i\pi)}{\varphi_2(-2u+8\eta)}\Big|_{u=\pm\theta_j},\label{hds2}
\eea
where $j=1,\cdots,N$ and the functions included in above operator identities are as follows
\bea\label{functions}
\hspace{-1.5truecm}\delta_1(u)=\hspace{-0.6truecm}&&-4(1-2e^{-\varepsilon}\sinh(u-\eta))(1+2e^{-\varepsilon}\sinh(u+\eta))(1-2e^{-\varepsilon^{\prime}}\sinh(u-\eta))\no\\[6pt]
&&\hspace{0.5truecm}\times(1+2e^{-\varepsilon^{\prime}}\sinh(u+\eta))\sinh(u+6\eta)\cosh(u+\eta)\sinh(u-6\eta)\no\\[6pt]
&&\hspace{0.5truecm}\times\cosh(u-\eta)\prod_{l=1}^{N}\varphi_1(u-\theta_l)\varphi_1(u+\theta_l),\no\\[6pt]
\hspace{-1.5truecm}\delta_2(u)=\hspace{-0.6truecm}&&-4(1-2e^{-\varepsilon}\sinh(u-\eta))(1-2e^{-\varepsilon^{\prime}}\sinh(u-\eta))\sinh(u+4\eta)\no\\[6pt]
&&\hspace{0.5truecm}\times\sinh(u-6\eta)\cosh^2(u-\eta)\prod_{l=1}^{N}\varphi_3(u-\theta_l)\varphi_3(u+\theta_l),\no\\[6pt]
\hspace{-1.5truecm}\varphi_2(u)=\hspace{-0.6truecm}&&-4\cosh(\frac{u}{2}-5\eta)\cosh(\frac{u}{2}-\eta)\sinh\frac{u}{2}\sinh(\frac{u}{2}-6\eta),\no\\[6pt]
\hspace{-1.5truecm}\varphi_3(u)=\hspace{-0.6truecm}&&-2\sinh(\frac{u}{2}+2\eta)\cosh(\frac{u}{2}-3\eta).
\eea
Moreover, the values of $t(u)$ at the points of $u=0,\,i\pi,\,6\eta,\,6\eta+i\pi$ can be calculated directly
\bea
&&t(0)=t(6\eta+i\pi)=(1+2e^{-\varepsilon}\sinh\eta)tr\{K^{+}(0)\}\prod_{l=1}^{N}\varphi_1(-\theta_l)\times {\rm id},\\
&&t(i\pi)=t(6\eta)=(1-2e^{-\varepsilon}\sinh\eta)tr\{K^{+}(i\pi)\}\prod_{l=1}^{N}\varphi_1(i\pi-\theta_l)\times {\rm id},
\eea
In addition, the transfer matrix of the model possesses the crossing symmetry and asymptotic behavior \cite{JHEP062014}
\bea
\hspace{-2truecm}&&t(u)=t(-u+6\eta+i\pi),\\[6pt]
\hspace{-2truecm}&&\lim_{u\rightarrow \pm\infty}t(u)=(\frac{1}{2})^{2N}e^{\mp 6(N+1)\eta-\varepsilon-\varepsilon^{\prime}}[1+2\cos(\varsigma^{\prime}-\varsigma-2i\eta)]\times{\rm id}+\cdots.
\eea
where the function $\varphi_1(u)$ is given by (\ref{p-1-function}).

Let $\Lambda(u)$ denote an eigenvalue of the transfer matrix $t(u)$. Then it follows from the above results on $t(u)$ that the eigenvalue $\Lambda(u)$ satisfies
\bea
\hspace{-2truecm}&&\Lambda(u)=\Lambda(-u+6\eta+i\pi),\label{Lam-crossing}\\[6pt]
&&\Lambda(\pm \theta_j)\Lambda(\pm \theta_j+6\eta+i\pi)=\frac{\delta_1(u)}{\varphi_1(2u)}\Big|_{u=\pm\theta_j},\quad j=1,\cdots,N,\label{Lam-hds1}\\[6pt]
&&\Lambda(\pm \theta_j)\Lambda(\pm \theta_j+4\eta)=\frac{\delta_2(u)\times \Lambda(u+2\eta+i\pi)}{\varphi_2(-2u+8\eta)}\Big|_{u=\pm\theta_j},\quad j=1,\cdots,N,\label{Lam-hds2}\\[6pt]
\hspace{-2truecm}&&\lim_{u\rightarrow \pm\infty}\Lambda(u)=(\frac{1}{2})^{2N}e^{\mp 6(N+1)\eta-\varepsilon-\varepsilon^{\prime}}[1+2\cos(\varsigma^{\prime}-\varsigma-2i\eta)]+\cdots,\label{asymptotic}\\[6pt]
&&\Lambda(0)=\Lambda(6\eta+i\pi)=(1+2e^{-\varepsilon}\sinh\eta)tr\{K^{+}(0)\}\prod_{l=1}^{N}\varphi_1(-\theta_l),\label{ts1}\\
&&\Lambda(i\pi)=\Lambda(6\eta)=(1-2e^{-\varepsilon}\sinh\eta)tr\{K^{+}(i\pi)\}\prod_{l=1}^{N}\varphi_1(i\pi-\theta_l),\label{ts2}
\eea
where the functions $\varphi_1(u)$, $\varphi_2(u)$, $\delta_1(u)$ and $\delta_2(u)$ are given by (\ref{p-1-function}) and (\ref{functions}). From these relations and the definition (\ref{trans}) of $t(u)$, we see that $\Lambda(u)$ is a trigonometric polynomial of $u$ with degree $4N+4$ and can be parameterized by its zeroes as
\bea
\Lambda(u)=\Lambda_0\prod_{j=1}^{2N+2}\sinh(\frac{u}{2}-\frac{z_j}{2}-\frac{3\eta}{2})\sinh(\frac{u}{2}+\frac{z_j}{2}-\frac{3\eta}{2}+\frac{i\pi}{2}),\label{Lamz}
\eea
where $\Lambda_0$ is a coefficient and $\{z_j|j=1,\cdots,2N+2\}$ are the zeroes of the polynomial. Putting the parameterization (\ref{Lamz}) into (\ref{Lam-crossing})-(\ref{ts2}), we can deduce the homogeneous BAEs
\bea
&&\hspace{-1.4truecm}\Lambda_0^2\prod_{l=1}^{2N+2}\sinh(\frac{\theta_j}{2}-\frac{z_l}{2}-\frac{3\eta}{2})\sinh(\frac{\theta_j}{2}+\frac{z_l}{2}-\frac{3\eta}{2}+\frac{i\pi}{2})
\sinh(\frac{\theta_j}{2}-\frac{z_l}{2}+\frac{3\eta}{2}+\frac{i\pi}{2})\no\\[6pt]
&&\hspace{-1truecm}\times\sinh(\frac{ \theta_j}{2}+\frac{z_l}{2}+\frac{3\eta}{2})=\frac{\delta_1(\theta_j)}{\varphi_1( 2\theta_j)},\label{BAEs-1}\\[6pt]
&&\hspace{-1.4truecm}\Lambda_0\prod_{l=1}^{2N+2}\sinh(\frac{\theta_j}{2}-\frac{z_l}{2}-\frac{3\eta}{2})\sinh(\frac{\theta_j}{2}+\frac{z_l}{2}-\frac{3\eta}{2}+\frac{i\pi}{2})
\sinh(\frac{\theta_j}{2}+\frac{z_l}{2}+\frac{\eta}{2}+\frac{i\pi}{2})\no\\[6pt]
&&\hspace{-1truecm}\times\sinh(\frac{ \theta_j}{2}-\frac{z_l}{2}+\frac{\eta}{2})=\frac{\delta_2(\theta_j)\prod_{l=1}^{2N+2}\sinh(\frac{ \theta_j}{2}-\frac{z_l}{2}+\frac{3\eta}{2})\sinh(\frac{ \theta_j}{2}+\frac{z_l}{2}+\frac{3\eta}{2})}{\varphi_2(-2\theta_j+8\eta)},\label{BAEs-2}
\eea
where $j=1,\cdots,N$. The BAEs and Eqs.(\ref{asymptotic})-(\ref{ts2}) can determine the $2N+3$ unknowns $\Lambda_0$ and $\{z_j|j=1,\cdots,2N+2\}$ completely.
Moreover, the energy spectrum of the system (\ref{IK-Ham}) can be expressed as
\bea
E=\frac{1}{2}\sum_{j=1}^{2N+2}[\coth(\frac{z_j}{2}+\frac{3\eta}{2})-\tanh(\frac{z_j}{2}-\frac{3\eta}{2})].\label{Energy-expressing}
\eea

\section{Patterns of zeroes}
\label{Patterns} \setcounter{equation}{0}

Now let us consider the patterns of the zeroes $\{z_j\}$ for the ground state. For convenience, we
put $\{\theta_j\equiv i\bar{\theta}_j\}$ with real $\{\bar{\theta}_j\}$, and let $\{\bar{z}_j\equiv iz_j\}$. In addition, we set $\chi_+={\rm arcsinh}(e^{\varepsilon}/2)$, $\chi_-={\rm arcsinh}(e^{\varepsilon^{\prime}}/2)$ and $\bar{\varsigma}^{\prime}=\varsigma^{\prime}-2i\eta$.
The hermiticity of the Hamiltonian (\ref{IK-Ham}) requires that the boundary parameters $\varsigma, \bar{\varsigma}^{\prime}\in (-\pi,\pi]$ and $\chi_+,\chi_-\in(0,+\infty)$.

\begin{figure}[htbp]
\centering
\includegraphics[scale=0.6]{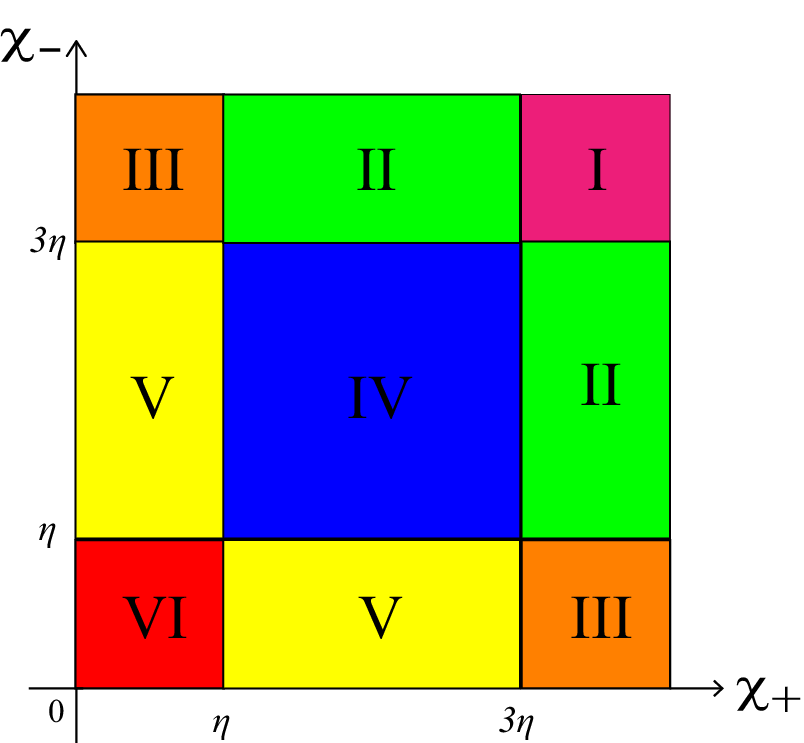}
  \caption{The distribution of $\bar{z}$-zeroes for the ground state in the $\chi_+$-$\chi_-$ plane.}\label{fig-region}
\end{figure}

By systematically varying the boundary parameters over their respective ranges with a fixed step, we obtain the solutions of Eqs.(\ref{BAEs-1})-(\ref{BAEs-2}) with $N=8$ and $\eta=0.35$. Summarizing the structures of solutions and carrying out the singularity analysis in the thermodynamic limit \cite{le}, the patterns of the $\bar{z}$-zeroes for the ground state include: (1) a set of bulk pairs $\{\bar{z}_j\sim \tilde{z}_j-5i\eta$, $(\tilde{z}_j-\pi)+5i\eta\}$ with real $\{\tilde{z}_j\}$ (the sign ``$\sim$" means approximately equal to due to the existence of finite size effect); (2) four free open boundary zeroes \footnote{The free open boundary zeroes indicate the zeroes that exist under open boundary conditions but are independent of boundary parameters.} $\pm\frac{\pi}{2}\pm 2i\eta$; (3) boundary pairs determined by the boundary parameters; (4) extra pairs with imaginary parts greater than $5\eta$. According to the number and type of the boundary pairs and additional zeroes, the distribution of the $\bar{z}$-zeroes can be divided into six different regimes in the $\chi_+$-$\chi_-$ plane, as shown in Fig.\ref{fig-region}.

\begin{figure}[htbp]
  \centering
\subfigure{\label{fig21:subfig:a}     \includegraphics[scale=0.63]{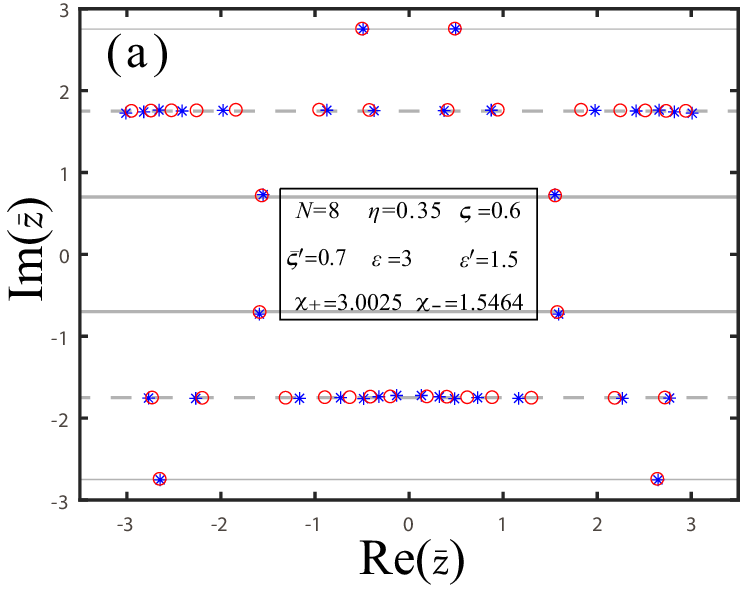}
    }  \subfigure{\label{fig21:subfig:b}   \includegraphics[scale=0.63]{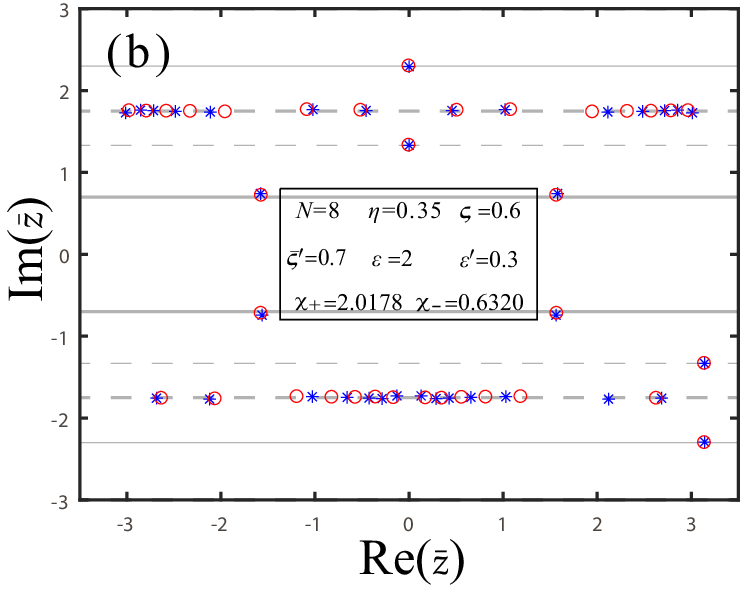}
    }
  \caption{Pattern of $\bar{z}$-zeroes  with certain model parameters for the ground state. (a) The boundary parameters are chosen in regime I. (b) The boundary parameters are chosen in regime II. The blue asterisks indicate $\bar{z}$-zeroes for $\{\bar{\theta}_j=0|j=1,\cdots,N\}$ and the red circles specify $\bar{z}$-zeroes
   with the inhomogeneity parameters $\{\bar{\theta}_j=0.1 j|j=1,\cdots,N\}$. The zeroes on the dashed thick lines, solid thick lines, dashed thin lines, and solid thin lines are classified as bulk pairs, free open boundary zeroes, boundary pairs, and extra pairs, respectively.}\label{fig41}
\end{figure}

1) In the regime I, where $\chi_+\geq 3\eta$, $\chi_-\geq 3\eta$, as shown in Fig.\ref{fig21:subfig:a}, all the $\bar{z}$-zeroes form $2N-2$ bulk pairs, four free open boundary zeroes and two extra pairs $\pm\alpha_1+ i\beta_1$, $\pm(\pi-\alpha_1)- i\beta_1$ with $\beta_1\geq5\eta$;

2) In the regime II, where $\chi_+\geq 3\eta$, $\eta\leq\chi_-< 3\eta$ or  $\eta\leq\chi_+< 3\eta$, $\chi_-\geq 3\eta$, as shown in Fig.\ref{fig21:subfig:b}, all the $\bar{z}$-zeroes form $2N-2$ bulk pairs, four free open boundary zeroes, one extra pair $i\beta_1$, $\pi-i\beta_1$ with $\beta_1\geq5\eta$, and one boundary pair $\pi-i(2\eta+\chi_1)$, $i(2\eta+\chi_1)$ with $\chi_{1}=\{\chi_+,\chi_-\}_{\min}$;

\begin{figure}[htbp]
  \centering
\subfigure{\label{fig32:subfig:a} \includegraphics[scale=0.63]{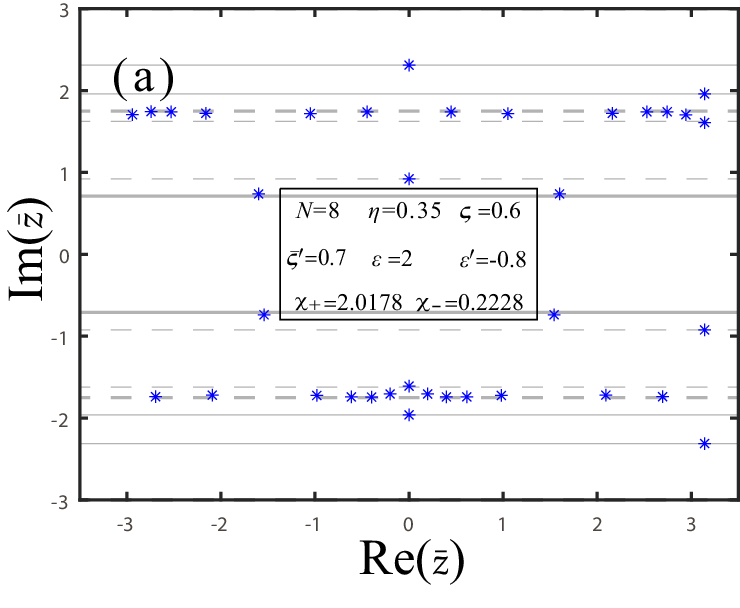} }\subfigure{\label{fig32:subfig:b} \includegraphics[scale=0.63]{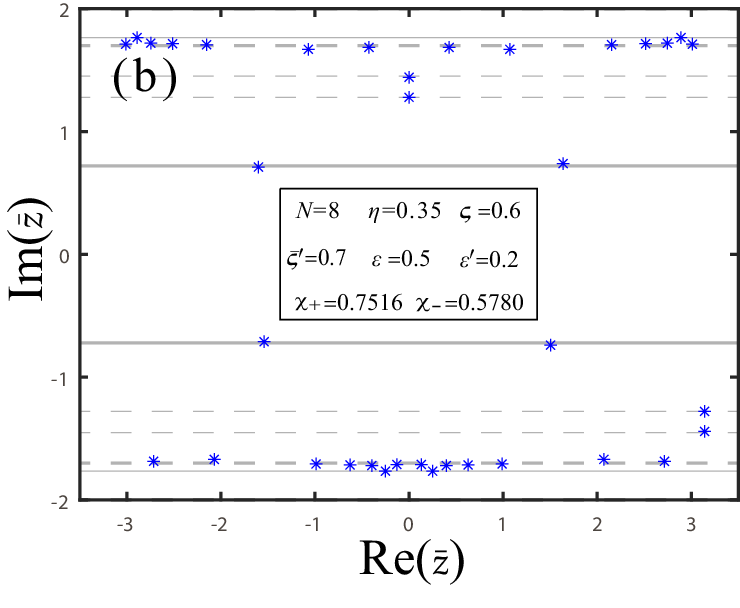}    } \subfigure{\label{fig32:subfig:c}\includegraphics[scale=0.63]{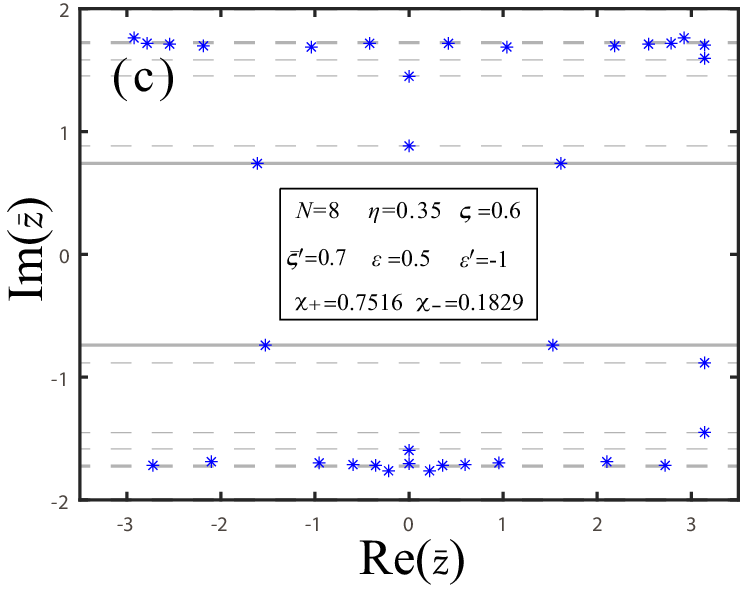}}
\subfigure{\label{fig32:subfig:d} \includegraphics[scale=0.63]{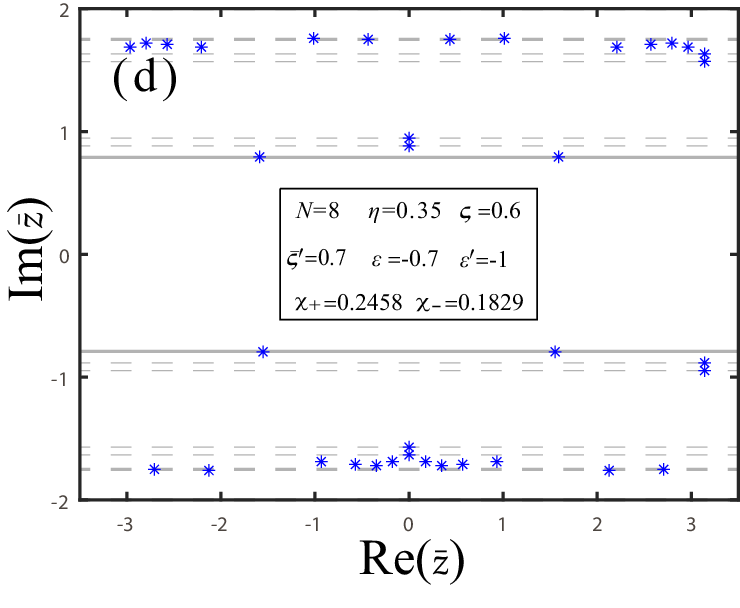}}
  \caption{Patterns of $\bar{z}$-zeroes with certain model parameters for $\{\bar{\theta}_j= 0|j=1,\cdots,N\}$ at the ground state. The boundary parameters in $(a)$-$(d)$ are chosen in the regimes III-VI, respectively. The zeroes on the dashed thick lines, solid thick lines, dashed thin lines, and solid thin lines are classified as bulk pairs, free open boundary zeroes, boundary pairs, and extra pairs, respectively.}\label{fig42}
\end{figure}

3) In the regime III, where $\chi_+\geq 3\eta$, $\chi_-< \eta$ or  $\chi_+< \eta$, $\chi_-\geq 3\eta$, as shown in Fig.\ref{fig32:subfig:a},
all the $\bar{z}$-zeroes form $2N-4$ bulk pairs, four free open boundary zeroes, two extra pairs $i\beta_1$, $-i\beta_2$, $\pi-i\beta_1$, $\pi+i\beta_2$ with $\{\beta_1,\beta_2\}\geq5\eta$, and two
boundary pairs $\pi-i(2\eta+\chi_1)$, $i(2\eta+\chi_1)$, $\pi+i(4\eta+\chi_1)$, $-i(4\eta+\chi_1)$ with $\chi_{1}=\{\chi_+,\chi_-\}_{\min}$;

4) In the regime IV, where $\eta\leq\chi_+< 3\eta$, $\eta\leq\chi_-< 3\eta$, as shown in Fig.\ref{fig32:subfig:b}, all the $\bar{z}$-zeroes form $2N-4$ bulk pairs, four free open boundary zeroes, two extra pairs $\pm\alpha_1-i\beta_1$, $\pm(\pi-\alpha_1)+i\beta_1$ with $\beta_1\geq5\eta$, and two
boundary pairs $\pi-i(2\eta+\chi_-)$,
 $i(2\eta+\chi_-)$, $\pi-i(2\eta+\chi_+)$,
 $i(2\eta+\chi_+)$;

5) In the regime V, where $\eta\leq\chi_+< 3\eta$, $\chi_-< \eta$ or  $\chi_+< \eta$, $\eta\leq\chi_-< 3\eta$, as shown in Fig.\ref{fig32:subfig:c}, all the $\bar{z}$-zeroes form $2N-4$ bulk pairs, four free open boundary zeroes, one extra pairs $-i\beta_1$, $\pi+i\beta_1$ with $\beta_1\geq5\eta$, and three boundary pairs $\pi-i(2\eta+\chi_-)$,
 $i(2\eta+\chi_-)$, $\pi-i(2\eta+\chi_+)$, $i(2\eta+\chi_+)$, $\pi+i(4\eta+\chi_1)$, $-i(4\eta+\chi_1)$ with $\chi_{1}=\{\chi_+,\chi_-\}_{\min}$;

6) In the regime VI, where $\chi_+<\eta$, $\chi_-<\eta$, as shown in Fig.\ref{fig32:subfig:d}, all the $\bar{z}$-zeroes form $2N-2$ bulk pairs, four free open boundary zeroes, and four boundary pairs $\pi-i(2\eta+\chi_-)$, $i(2\eta+\chi_-)$, $\pi-i(2\eta+\chi_+)$, $i(2\eta+\chi_+)$, $\pi+i(4\eta+\chi_-)$, $-i(4\eta+\chi_-)$ and $\pi+i(4\eta+\chi_+)$, $-i(4\eta+\chi_+)$.
 
We find that the patterns of zeroes for the IK model are different from the ones for models related to untwisted $A$-type algebras \cite{PRB1022020,PRBL1032021,NPB9752022,NPB9842022}.
In Tab.\ref{open-spin1-IK-table}, we list the differences between the zeroes of spin-1 model (the 19-vertex model related to untwisted $A^{(1)}_1$) and the IK model. We observe that the zeroes of the spin-1 model form conjugate pairs and the bulk pairs are a set of four strings. However, the zeroes in the IK model form non-conjugated pairs and the bulk pairs are located only on the two lines with imaginary parts close to $\pm 5\eta$.

\begin{table}\renewcommand\arraystretch{2.1}\caption{Patterns of zeroes distribution at the ground state for the spin-1 model and the IK model. Here, $\lambda$ is a real number and $z_x$ is a real number larger than $\frac{3\eta}{2}$. The parameters $z_1, z_2$ are two real numbers that tend to infinity in the thermodynamic limit.}

\resizebox{\textwidth}{!}{
\begin{tabular}{|c|c|c|}
\hline &  {\bf Zeroes of the spin-1 model} & {\bf Zeroes of the IK model}\\

\hline $\mathrm{1}$ & Conjugated bulk pairs $\{\bar{z}^{(1)}_k\sim \tilde{z}^{(1)}_k\pm ni\,|\,n=1,2\}$ & Non-conjugated bulk pairs $\{\bar{z}_j\sim \tilde{z}_j-5i\eta$, $(\tilde{z}_j-\pi)+5i\eta\}$\\

\hline $\mathrm{2}$ & Real free open boundary zeroes: $\pm 0$ & Imaginary free open boundary zeroes: $\pm\frac{\pi}{2}\pm 2i\eta$ \\

\hline $\mathrm{3}$ & \makecell{Conjugated boundary pairs: $\pm[{\rm min}(p,q)]i,\pm[{\rm max}(p,q)]i$,\\ $\pm[1\pm{\rm min}(p,q)]i,\pm[1+{\rm max}(p,q)]i$,\\$\pm[\frac{3}{2}\pm{\rm min}(p,q)]i$} & \makecell{Non-conjugated boundary pairs: $\pi-i(2\eta+\chi_-)$, $i(2\eta+\chi_-)$, $\pi-i(2\eta+\chi_+)$,\\$i(2\eta+\chi_+)$, $\pi+i(4\eta+\chi_-)$, $-i(4\eta+\chi_-)$,\\$\pi+i(4\eta+\chi_+)$, $-i(4\eta+\chi_+)$} \\
\hline

$\mathrm{4}$ & \makecell{Conjugated extra pairs: $\pm z_1,\pm z_2,\pm\left(z_x-\frac{1}{2}\right) i$,\\$\pm\left(z_x+\frac{1}{2}\right) i,\pm(\lambda\pm\frac{3i}{2})$} & \makecell{Non-conjugated extra pairs: $\pm\alpha_1+ i\beta_1$, $\pm(\pi-\alpha_1)- i\beta_1$ ,$i\beta_1$, \\$-i\beta_2$, $\pi-i\beta_1$, $\pi+i\beta_2$} \\
\hline
\end{tabular}}

\label{open-spin1-IK-table}
\end{table}

\section{Surface energy}\label{Interaction-case1}
\setcounter{equation}{0}

We now study the surface energy induced by the generic boundary fields. The surface energy is defined by $E_b=E_g-E_p$, where $E_g$ is the ground state energy of Hamiltonian (\ref{IK-Ham})
and $E_p$ is the ground state energy of the periodic IK model. In the thermodynamic limit, the distribution of $\tilde{z}$-zeroes can be described by the continuum density $\rho(\tilde{z})$. Furthermore, we assume that the inhomogeneity parameters also have the continuum density
$\sigma(\bar{\theta})=1/[2N(\bar{\theta}_j-\bar{\theta}_{j-1} )]$.

In regime I, taking the logarithm of the BAEs (\ref{BAEs-1}) or (\ref{BAEs-2}) and making the difference of $\bar{\theta}_{j-1}$ and $\bar{\theta}_j$ in the thermodynamic limit, we readily have
\bea
&&2N\int_{-\pi}^{\pi}[b_2(u\hspace{-0.06truecm}-\hspace{-0.06truecm}\bar{z})\hspace{-0.06truecm}+\hspace{-0.06truecm}b_{8}(u\hspace{-0.06truecm}-\hspace{-0.06truecm}\bar{z}\hspace{-0.06truecm}+\hspace{-0.06truecm}\pi)]\rho(\bar{z})d\bar{z}\hspace{-0.06truecm}+\hspace{-0.06truecm}b_5(u\hspace{-0.06truecm}+\hspace{-0.06truecm}\frac{\pi}{2})\hspace{-0.06truecm}+\hspace{-0.06truecm}b_5(u\hspace{-0.06truecm}-\hspace{-0.06truecm}\frac{\pi}{2})\no\\[6pt]
&&\qquad+b_{\frac{\beta_1}{\eta}+3}(u\hspace{-0.06truecm}-\hspace{-0.06truecm}\alpha_1)\hspace{-0.06truecm}+\hspace{-0.06truecm}b_{\frac{\beta_1}{\eta}-3}(u+\pi-\alpha_1)\hspace{-0.06truecm}+\hspace{-0.06truecm}b_{\frac{\beta_1}{\eta}-3}(u-\pi+\alpha_1)+b_{\frac{\beta_1}{\eta}+3}(u\hspace{-0.06truecm}+\hspace{-0.06truecm}\alpha_1)\no\\[6pt]
=\hspace{-0.6truecm}&&b_{|1+\frac{\chi_+}{\eta}|}(u)+b_{|1-\frac{\chi_+}{\eta}|}(u\hspace{-0.06truecm}+\hspace{-0.06truecm}\pi)\hspace{-0.06truecm}+\hspace{-0.06truecm}b_{|1+\frac{\chi_-}{\eta}|}(u)+b_{|1-\frac{\chi_-}{\eta}|}(u\hspace{-0.06truecm}+\hspace{-0.06truecm}\pi)\hspace{-0.06truecm}+\hspace{-0.06truecm}b_6(u)\hspace{-0.06truecm}+\hspace{-0.06truecm}b_6(u\hspace{-0.06truecm}+\hspace{-0.06truecm}\pi)\hspace{-0.06truecm}-\hspace{-0.06truecm}b_2(u)
\no\\[6pt]&&-b_2(u\hspace{-0.06truecm}+\hspace{-0.06truecm}\pi)\hspace{-0.06truecm}-\hspace{-0.06truecm}b_3(u\hspace{-0.06truecm}-\hspace{-0.06truecm}\frac{\pi}{2})\hspace{-0.06truecm}-\hspace{-0.06truecm}b_3(u\hspace{-0.06truecm}+\hspace{-0.06truecm}\frac{\pi}{2})
\hspace{-0.06truecm}+\hspace{-0.06truecm}2N\int_{-\pi}^{\pi}[b_4(u\hspace{-0.06truecm}-\hspace{-0.06truecm}\bar{\theta})+b_{6}(u\hspace{-0.06truecm}-\hspace{-0.06truecm}\bar{\theta}\hspace{-0.06truecm}+\hspace{-0.06truecm}\pi)]\sigma(\bar{\theta})d\bar{\theta},\label{int-p}
\eea
where the function $b_n(u)=\frac{1}{2\pi}\frac{\sin(u)}{\cosh(n\eta)-\cos u}$. Solving the Eq.(\ref{int-p}) by Fourier transformation, we obtain
\bea
\hspace{-0.35truecm}\tilde{\rho}(k)=
&&\hspace{-0.6truecm}[2N(\tilde{b}_4+(-1)^k\tilde{b}_6)\tilde{\sigma}(k)+\tilde{b}_{|1+\frac{\chi_+}{\eta}|}+\tilde{b}_{|1+\frac{\chi_-}{\eta}|}+(-1)^k\cdot(\tilde{b}_{|1-\frac{\chi_+}{\eta}|}+\tilde{b}_{|1-\frac{\chi_-}{\eta}|}+\tilde{b}_{6}-\tilde{b}_{2})\no\\[6pt]
&&\hspace{-0.8truecm}+\tilde{b}_{6}-\tilde{b}_{2}-2(\tilde{b}_3+\tilde{b}_5 ) \cos(\frac{\pi k}{2})-2(\tilde{b}_{\frac{\beta_1}{\eta}+3}+\tilde{b}_{\frac{\beta_1}{\eta}-3})\cos(\alpha_1 k)]/[2N(\tilde{b}_2+(-1)^k\tilde{b}_8)],\label{pro1}
\eea
where the Fourier spectrum $k$ takes integer values and $\tilde{b}_n(k)=sign(k)ie^{-\eta|nk|}$. In the homogeneous limit $\sigma(\bar{\theta})=\delta(\bar{\theta})$, the ground state energy $E_{g1}$ of the
Hamiltonian (\ref{IK-Ham}) in regime I can be expressed as (the proof is relegated to appendix A)
\bea
E_{g1}=&&\hspace{-0.6truecm}N\sum_{k=-\infty}^{\infty}[\tilde{a}_8(k)e^{-i\pi k}-\tilde{a}_2(k)]\tilde{\rho}(k)+\frac{1}{2}\Big[\coth(\frac{3\eta+i\alpha_1-\beta_1}{2})+\coth(\frac{3\eta-i\alpha_1-\beta_1}{2})\no\\
&&\hspace{-0.6truecm}+\tanh(\frac{3\eta-i\alpha_1+\beta_1}{2})+\tanh(\frac{3\eta+i\alpha_1+\beta_1}{2})\Big]+\tanh \eta+\tanh(5\eta),\label{Eg1}
\eea
where $\tilde{a}_n(k)=e^{-\eta|nk|}$ is the Fourier transformation of $a_n(u)=\frac{1}{2\pi}\frac{\sinh(n\eta)}{\cosh(n\eta)-\cos u}$. The ground state energy of the periodic IK model is (the proof is relegated to appendix B)
\bea
E_p=2N\sum_{k=1}^{\infty}[(-1)^k e^{-\eta|6k|}-1]\frac{e^{-\eta|4k|}+(-1)^k e^{-\eta|6k|} }{1+(-1)^ke^{-\eta|6k|}}.\label{Ep}
\eea
Subtracting Eq.(\ref{Ep}) from Eq.(\ref{Eg1}), we obtain the surface energy in the regime I
\bea
\hspace{-1truecm}&&E_{b1}=e_b(\chi_+)+e_b(\chi_-)+e_{b0},\label{Eb1}\\[8pt]
\hspace{-1truecm}&&e_b(\chi)=\sum_{k=1}^{\infty}[(-1)^k e^{-\eta|6k|}-1]\frac{e^{-\eta|(1+\frac{\chi}{\eta})k|}+(-1)^k e^{-\eta|(1-\frac{\chi}{\eta})k|} }{1+(-1)^ke^{-\eta|6k|}},\\[6pt]
\hspace{-1truecm}&&e_{b0}=\sum_{k=1}^{\infty}[(-1)^k e^{-\eta|6k|}-1]\frac{(e^{-\eta|6k|}-e^{-\eta|2k|})(1+(-1)^k)-2(e^{-\eta|5k|}+e^{-\eta|3k|})\cos\frac{\pi k}{2} }{1+(-1)^k e^{-\eta|6k|}}\no\no\\[6pt]
&&\hspace{1.1truecm} +\tanh \eta+\tanh(5\eta),
\eea
where $e_b(\chi)$ indicates the contribution of one boundary field and $e_{b0}$ is the surface energy induced by the free open boundary. The absence of $\alpha_1$ and $\beta_1$ in Eq.(\ref{Eb1}) is due to that the bare contribution of the additional zeroes to the energy is exactly offset by that of the change of the continuous zero density
\bea\label{add0}
&&\hspace{-0.6truecm}\frac{1}{2}\Big[\coth(\frac{3\eta\hspace{-0.06truecm}+\hspace{-0.06truecm}i\alpha_1\hspace{-0.06truecm}-\hspace{-0.06truecm}\beta_1}{2})\hspace{-0.06truecm}+\hspace{-0.06truecm}\coth(\frac{3\eta\hspace{-0.06truecm}-\hspace{-0.06truecm}i\alpha_1\hspace{-0.06truecm}-\hspace{-0.06truecm}\beta_1}{2})\hspace{-0.06truecm}+\hspace{-0.06truecm}\tanh(\frac{3\eta\hspace{-0.06truecm}-\hspace{-0.06truecm}i\alpha_1\hspace{-0.06truecm}+\hspace{-0.06truecm}\beta_1}{2})\hspace{-0.06truecm}+\hspace{-0.06truecm}\tanh(\frac{3\eta\hspace{-0.06truecm}+\hspace{-0.06truecm}i\alpha_1\hspace{-0.06truecm}+\hspace{-0.06truecm}\beta_1}{2})\Big]
\no\\[6pt]
&&\hspace{-0.2truecm}-\hspace{-0.06truecm}\sum_{k=-\infty}^{\infty}[(-1)^k e^{-\eta|6k|}\hspace{-0.06truecm}-\hspace{-0.06truecm}1]\frac{(e^{-|(\beta_1-3\eta)k|}\hspace{-0.06truecm}+\hspace{-0.06truecm}e^{-|(\beta_1+3\eta)k|})\cos(\alpha_1 k)}{1\hspace{-0.06truecm}+\hspace{-0.06truecm}(-1)^k e^{-\eta|6k|}}\hspace{-0.06truecm}=\hspace{-0.06truecm}0.
\eea
For the regimes II and IV, we use the similar procedure to that applied in regime I and find that the boundary pairs and additional zeroes will also contribute nothing to the surface energy. The surface energy $E_{b2}$ and $E_{b4}$ take exactly the same form as (\ref{Eb1}).

In regime III, the imaginary part of the inner boundary pair will satisfy $\{\chi_+,\chi_-\}_{\min}+2\eta<3\eta$. In this case, the inner boundary pair indeed contributes a nonzero value to energy and the surface energy reads
\bea
E_{b3}=&&\hspace{-0.6truecm}\sum_{k=1}^{\infty}[1-(-1)^k e^{-\eta|6k|}]\frac{(-1)^k e^{-\eta|(1-\frac{\chi_1}{\eta})k|}+e^{-\eta|(5+\frac{\chi_1}{\eta})k|} }{1+(-1)^k e^{-\eta|6k|}}+E_{b1}\no\\
&&\hspace{-0.6truecm}+\frac{1}{2}\coth(\frac{5\eta}{2}+\frac{\chi_1}{2})+\frac{1}{2}\tanh(\frac{\eta}{2}-\frac{\chi_1}{2}),
\eea
where $\chi_{1}=\{\chi_+,\chi_-\}_{\min}$. The contributions of the two boundaries to the surface energy are no longer additive, which implies that a correlational effect appears between the two boundary fields in this case.

In the regimes V and VI, there will exist another inner boundary pair that satisfies $\{\chi_+,\chi_-\}_{\max}+2\eta<3\eta$. The new inner boundary pair also contributes a nonzero value to energy and the surface energy reads
\bea
E_{b5}=E_{b6}=&&\hspace{-0.6truecm}\sum_{k=1}^{\infty}[1-(-1)^k e^{-\eta|6k|}]\frac{(-1)^k e^{-\eta|(1-\frac{\chi_2}{\eta})k|}+e^{-\eta|(5+\frac{\chi_2}{\eta})k|} }{1+(-1)^k e^{-\eta|6k|}}+E_{b3}\no\\
&&\hspace{-0.6truecm}+\frac{1}{2}\coth(\frac{5\eta}{2}+\frac{\chi_2}{2})+\frac{1}{2}\tanh(\frac{\eta}{2}-\frac{\chi_2}{2}),
\eea
where $\chi_{2}=\{\chi_+,\chi_-\}_{\max}$. In the two regimes, a correlation effect also appears between the two boundary fields.

We note that the boundary parameters $\varsigma$ and $\varsigma^{\prime}$ do not appear in the surface energies of all regimes, implying
that they contribute nothing to the surface energy in the leading order.
The surface energies with different boundary parameters $\varepsilon$ and $\varepsilon^{\prime}$ are shown in Fig.\ref{fig-surfaceE} below. From this figure, we observe that the surface energies with the increasing or decreasing of $\varepsilon$ tend to fixed values. In addition, the surface energy behaves as a monotonic function of $\varepsilon$ or $\varepsilon^{\prime}$ when $\chi_+<\eta$ or $\chi_-<\eta$.

\begin{figure}[htp]
\centering
\includegraphics[width=12cm]{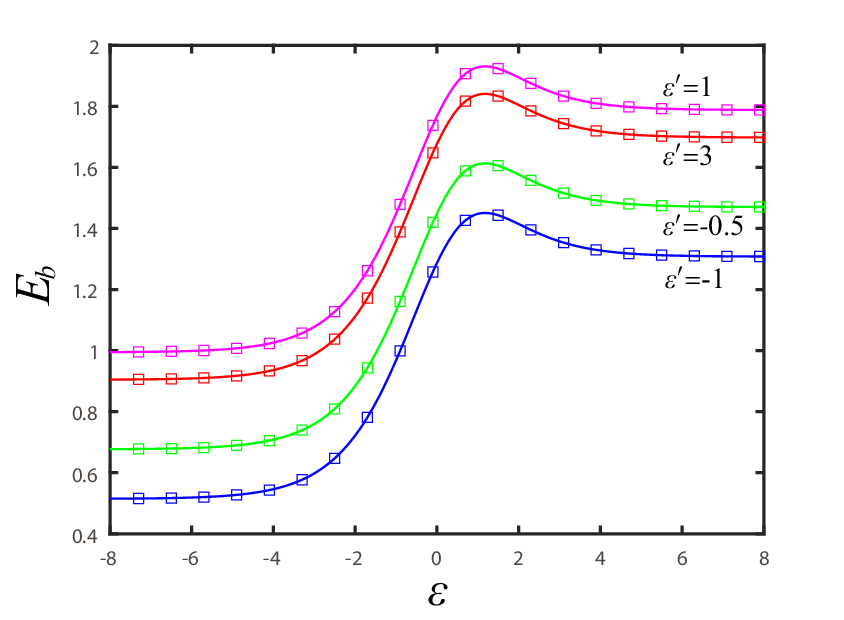}
  \caption{Surface energies versus the boundary parameters $\varepsilon$ and $\varepsilon^{\prime}$ for $\eta=0.5$. The lines indicate the analytic results and the squares indicate the DMRG results for $N=180$.}
\label{fig-surfaceE}
\end{figure}

\section{Boundary excitations}
\label{B-excitation} \setcounter{equation}{0}
Besides the usual bulk elementary excitations, the system also exists the boundary excitations associated with the boundary magnetic fields (\ref{LRboundary}). By comparing the zeroes distributions of the ground state and the excited states, we find that the boundary excitations can exist in the regimes II-VI, where the boundary parameter $\chi_+< 3\eta$ or  $\chi_-< 3\eta$. The typical boundary excitation is putting the boundary pairs from $\pi-i(2\eta+\chi_{\pm})$, $i(2\eta+\chi_{\pm})$ to $\pi-i(4\eta-\chi_{\pm})$, $i(4\eta-\chi_{\pm})$, and from $\pi+i(4\eta+\chi_{\pm})$, $-i(4\eta+\chi_{\pm})$ (or the bulk pair) to $\pi+i(2\eta-\chi_{\pm})$, $-i(2\eta-\chi_{\pm})$. These four new boundary pairs indeed satisfy the BAEs (\ref{BAEs-1})-(\ref{BAEs-2}).

\begin{figure}[htp]
\centering
\subfigure{\label{5a}
\includegraphics[scale=0.52]{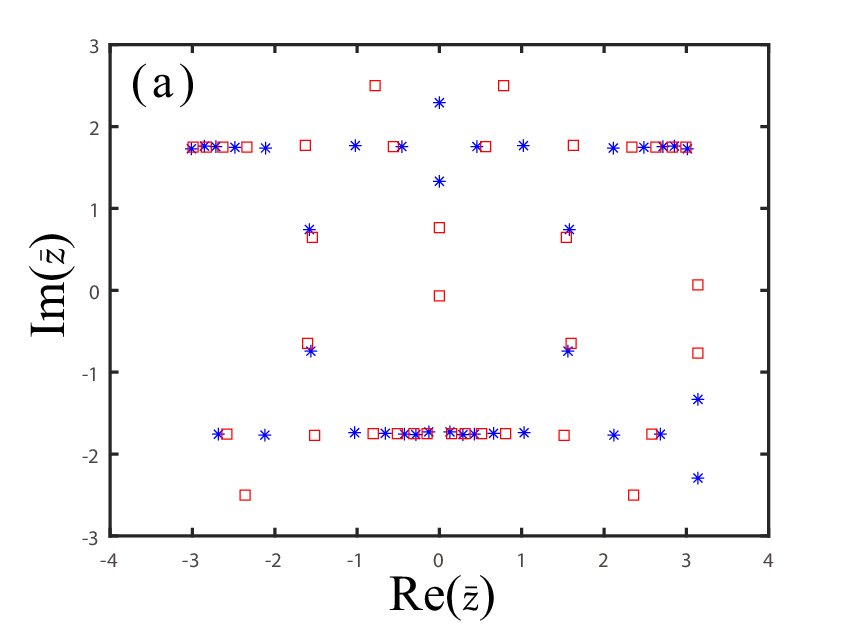}
}
\subfigure{\label{5b}
\includegraphics[scale=0.52]{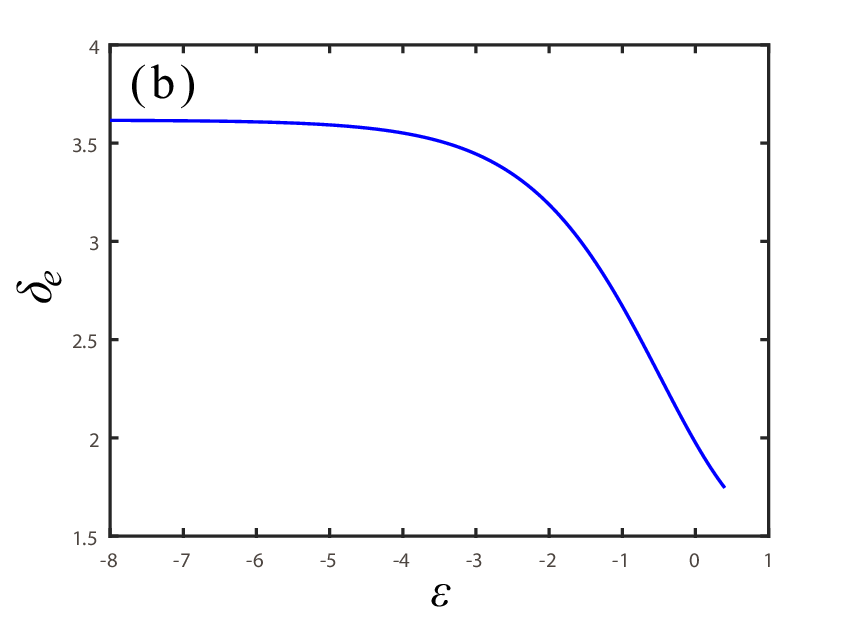}
}
\caption{(a) The distribution of $\bar{z}$-zeroes for $\{\bar{\theta}_j= 0|j=1,\cdots,N\}$   with $N=8$, $\eta=0.35$, $\varsigma=0.6$, $\bar{\varsigma}^{\prime}=0.7$, $\varepsilon=2$ and $\varepsilon^{\prime}=0.3$. Here the blue asterisks represent the pattern of zeroes at the ground state and the red squares denote those at the 32nd excited state with boundary pairs $\pi+i(2\eta-\chi_{-})$, $-i(2\eta-\chi_{-})$, $\pi-i(4\eta-\chi_{-})$, $i(4\eta-\chi_{-})$.
(b) The excitation energy versus $\varepsilon$ in the thermodynamic limit.}\label{fig5}
\end{figure}

As an example, we show in Fig.\ref{5a} the pattern of zeroes at the ground state (blue asterisks) and that at the 32nd excited state (red squares) with two boundary pairs $\pi+i(2\eta-\chi_{-})$, $-i(2\eta-\chi_{-})$, $\pi-i(4\eta-\chi_{-})$, $i(4\eta-\chi_{-})$ in regime II with $N=8$. In the excitation process, we observe that four zeroes at the bulk pair parts of the ground state jump to form $\pm\alpha_1+ i\beta_1$, $\pm(\pi-\alpha_1)- i\beta_1$ with $\beta_1>5\eta$. From the Eq.(\ref{add0}), we have known that the change of the zeroes contribute nothing to the energy. Without losing generality, we omit the zeroes in the subsequent discussions. In the thermodynamic limit, the resulted density change between ground and excited states reads
\bea
\delta\tilde{\rho}_{\chi_-}(k)=\frac{e^{-\eta|(5+\frac{\chi_-}{\eta})k|}-e^{-\eta|(7-\frac{\chi_-}{\eta})k|}-e^{-\eta|(1+\frac{\chi_-}{\eta})k|}-(-1)^k e^{-\eta|(5-\frac{\chi_-}{\eta})k|} }{2N(e^{-\eta|2k|}+(-1)^k e^{-\eta|8k|})}.
\eea
The corresponding excited energy is
\bea
\delta_{e_{\chi_-}}=&&\hspace{-0.6truecm}\sum_{k=1}^{\infty}[(-1)^k e^{-\eta|6k|}-1]\frac{e^{-\eta|(5+\frac{\chi_-}{\eta})k|}-e^{-\eta|(7-\frac{\chi_-}{\eta})k|}-e^{-\eta|(1+\frac{\chi_-}{\eta})k|}-(-1)^k e^{-\eta|(5-\frac{\chi_-}{\eta})k|}  }{1+(-1)^k e^{-\eta|6k|}}\no\\[6pt]&&\hspace{-0.6truecm}+\frac{1}{2}\coth(\frac{7\eta}{2}-\frac{\chi_-}{2})+\frac{1}{2}\coth(\frac{\eta}{2}+\frac{\chi_-}{2})+\frac{1}{2}\tanh(\frac{5\eta}{2}-\frac{\chi_-}{2})\no\\[6pt]&&\hspace{-0.6truecm}-\frac{1}{2}\coth(\frac{5\eta}{2}+\frac{\chi_-}{2})-\tanh(\frac{\eta}{2}-\frac{\chi_-}{2}).\label{be}
\eea
The excited energies versus $\varepsilon$ are shown in Fig.\ref{5b}.
Using the similar idea and after tedious calculations, we obtain the boundary excitations in the
regimes II-VI
\bea
\delta_{e}=
\begin{cases}
\delta_{e_{\chi_+}} {\rm or\hspace{4pt}} \delta_{e_{\chi_-}},\quad {\rm  in\hspace{4pt} regime }\hspace{4pt} {\rm II\hspace{4pt} and\hspace{4pt} III}, \\
\delta_{e_{\chi_+}} {\rm or\hspace{4pt}} \delta_{e_{\chi_-}} {\rm or\hspace{4pt}} \delta_{e_{\chi_+}}+\delta_{e_{\chi_-}},\quad {\rm  in\hspace{4pt} regime }\hspace{4pt} {\rm IV\hspace{4pt}, V\hspace{4pt} and\hspace{4pt} VI}.
\end{cases}
\eea


\section{Conclusions}
\label{Con}
In this paper, we have extended the $t$-$W$ method to the IK model with generic integrable boundaries. By parameterizing the eigenvalue of the transfer matrix using its zeroes, we derive two sets of homogeneous zeroes BAEs. The patterns of zeroes distributions in different regimes are obtained by solving the BAEs. Unlike untwisted $A$-type models, the zeroes of the present system no longer form conjugate pairs. Based on the patterns, we have obtained the exact surface energies and boundary excitations in different regimes of the boundary parameters. The results of the surface energy indicate that a correlation effect appears between the two boundary fields in some regimes. From the analytic results and the DMRG results, we also find that the boundary parameters $\varsigma$ and $\varsigma^{\prime}$ contribute nothing to the surface energy in the leading order $o(1/N)$. This is due to the fact that the parameters $\varsigma$ and $\varsigma^{\prime}$ can be considered as the twisted angle between two unparallel boundary magnetic fields and this twisted angle does not contribute to the $o(1/N)$ order of the energy in the thermodynamic limit. However, the parameters $\varsigma$ and $\varsigma^{\prime}$ are useful for studying other physical properties. For example, the unparallel boundary magnetic fields can give rise to the helical states, which can be calculated based on the obtained eigenvalues. Through these helical states, we can further study some persistent currents that exist within the system. In addition, based on the patterns of zeroes, other physical quantities such as the free energy at finite temperature, specific heat and magnetic susceptibility in the external magnetic field can be studied. It is also important to combine the quantum transfer matrix method \cite{JSM2017023106,
JSM2018113102} and the $t$-$W$ method to explore the quench dynamics of the system.

The method and process presented in this paper can be generalized to other models with twisted affine algebra symmetries. Explicit $R$-matrices for models related to quantum twisted affine algebras can be constructed by the general procedure developed in \cite{DGZ96}. Work on the quantum integrable
$D_2^{(2)}$ spin chain with generic boundary fields \cite{JHEP2022101} is in progress and results will be presented elsewhere.

\section*{Acknowledgments}

We thank Professor Yupeng Wang for valuable discussions. We acknowledge the financial support from Australian Research Council Discovery Project DP190101529, Future Fellowship FT180100099, National Key R$\&$D Program of China (Grant No.2021YFA1402104),
China Postdoctoral Science Foundation Fellowship 2020M680724, National Natural Science Foundation of China (Grant Nos. 12074410, 12434006, 12247103, 12247179, 11934015 and 11975183), the Major Basic Research Program of Natural Science of Shaanxi Province (Grant Nos. 2021JCW-19 and 2017ZDJC-32), and the Strategic Priority Research Program of the Chinese Academy of Sciences (Grant No. XDB33000000).

\appendix
\section{Proof of the ground state energy (\ref{Eg1}) in regime I}
\setcounter{equation}{0}
\renewcommand{\theequation}{A.\arabic{equation}}
In regime I, all the $\bar{z}$-zeroes for the ground state form $2N-2$ bulk pairs $\{\bar{z}_j\sim \tilde{z}_j-5i\eta$, $(\tilde{z}_j-\pi)+5i\eta|j=1,\cdots,2N-2\}$ with real $\{\tilde{z}_j\}$, four free open boundary zeroes $\pm\frac{\pi}{2}\pm 2i\eta$ and two extra pairs $\pm\alpha_1+ i\beta_1$, $\pm(\pi-\alpha_1)- i\beta_1$ with $\beta_1\geq5\eta$. Substituting the  patterns into the expression (\ref{Energy-expressing}) of the energy spectrum and using $\{\bar{z}_j\equiv iz_j\}$, we have
\bea
E_{g1}=&&\hspace{-0.6truecm}\frac{i}{2}\sum_{j=1}^{2N-2}[\cot(\frac{\tilde{z}_j}{2}-i\eta)+\tan(\frac{\tilde{z}_j}{2}-4i\eta)]+\frac{1}{2}\Big[\coth(\frac{3\eta+i\alpha_1-\beta_1}{2})+\coth(\frac{3\eta-i\alpha_1-\beta_1}{2})\no\\
&&\hspace{-0.6truecm}+\tanh(\frac{3\eta-i\alpha_1+\beta_1}{2})+\tanh(\frac{3\eta+i\alpha_1+\beta_1}{2})\Big]+\tanh \eta+\tanh(5\eta).\label{Eg1-proof1}
\eea
The hermiticity of the Hamiltonian (\ref{IK-Ham}) requires $\{\tilde{z}_j\}$ to be symmetric along the imaginary axis. Therefore the first term on the RHS of (\ref{Eg1-proof1}) can be expressed as
\bea
&&\hspace{-0.6truecm}\frac{i}{2}\sum_{j=1}^{2N-2}[\cot(\frac{\tilde{z}_j}{2}-i\eta)+\tan(\frac{\tilde{z}_j}{2}-4i\eta)]\no\\[6pt]
=&&\hspace{-0.6truecm}\frac{i}{4}\sum_{j=1}^{2N-2}[\cot(\frac{\tilde{z}_j}{2}-i\eta)-\cot(\frac{\tilde{z}_j}{2}+i\eta)+\tan(\frac{\tilde{z}_j}{2}-4i\eta)-\tan(\frac{\tilde{z}_j}{2}+4i\eta)]\no\\[6pt]
=&&\hspace{-0.6truecm}-\frac{1}{2}\sum_{j=1}^{2N-2}[\frac{\sinh (2\eta)}{\cosh (2\eta)-\cos (\tilde{z}_j)}-\frac{\sinh (8\eta)}{\cosh (8\eta)-\cos (\tilde{z}_j+\pi)}]\no\\
=&&\hspace{-0.6truecm} \pi \sum_{j=1}^{2N-2} [a_8(\tilde{z}_j+\pi)-a_2(\tilde{z}_j) ]\no\\
=&&\hspace{-0.6truecm}\ldots,\no
\eea
In the thermodynamic limit, the distribution of $\tilde{z}$-zeroes can be  described by the continuum density $\rho(\tilde{z})$
\bea
\ldots=2\pi N \int_{-\pi}^{\pi}[a_8(\tilde{z}+\pi)-a_2(\tilde{z}) ]\rho(\tilde{z})d\tilde{z}=\ldots,\no
\eea
then using the Fourier transformation one obtain
\bea
\ldots=N\sum_{k=-\infty}^{\infty}[\tilde{a}_8(k)e^{-i\pi k}-\tilde{a}_2(k)]\tilde{\rho}(k).\label{Eg1-proof2}
\eea
Finally, replacing the first term on the LHS of (\ref{Eg1-proof1}) with (\ref{Eg1-proof2}), the expression (\ref{Eg1}) is achieved. This completes our proof.

\section{Proof of the ground state energy for the periodic IK model}
\setcounter{equation}{0}
\renewcommand{\theequation}{B.\arabic{equation}}
For the IK model with periodic boundary condition, the $t$-$W$ process is natural and simple. In this case, the transfer matrix  read
\bea
t^p(u)=tr_0 T_0(u),
\eea
and the following operator identities can be obtained
\bea
&&t^p(\theta_j)t^p(\theta_j+6\eta+i\pi)=\tilde{a}(\theta_j)\tilde{d}(\theta_j+6\eta+i\pi)\times {\rm id},\\
&&t^p(\theta_j)t^p(\theta_j+4\eta)=\tilde{\delta}(\theta_j)t^p(\theta_j+2\eta+i\pi),\quad j=1,\ldots,N,
\eea
where
\bea
&&\tilde{a}(u)=\prod_{l=1}^{N}h_3(u-\theta_l),\quad \tilde{d}(u)=\prod_{l=1}^{N}h_4(u-\theta_l),\\
&&\tilde{\delta}(u)=(-2)^N\prod_{l=1}^N\cosh\Big(\frac{u-\theta_l}{2}-3\eta\Big)\sinh\Big(\frac{u-\theta_l}{2}+2\eta\Big).
\eea
Here the functions $h_3(u)$ and $h_4(u)$ are given by (\ref{R-element-2}). In addition, the asymptotic behavior of $t^p(u)$ for $u\rightarrow \pm\infty$ reads
\bea
\lim_{u\rightarrow \pm\infty}t^p(u)=(\frac{1}{2})^Ne^{\mp 3N\eta}[1+2\cosh(2\hat{M}\eta)]\times{\rm id}+\cdots,
\eea
where $\hat{M}$ is related to $U(1)$ charge operator $\hat{M}=\sum_{j=1}^N(E_j^{11}-E_j^{33})$.

Let $\Lambda^p(u)$ denote an eigenvalue of the transfer matrix $t^p(u)$. Then it follows from the above results on $t^p(u)$ that the eigenvalue $\Lambda^p(u)$ satisfies
\bea
&&\Lambda^p(\theta_j)\Lambda^p(\theta_j+6\eta+i\pi)=\tilde{a}(\theta_j)\tilde{d}(\theta_j+6\eta+i\pi),\label{tsp1}\\
&&\Lambda^p(\theta_j)\Lambda^p(\theta_j+4\eta)=\tilde{\delta}(\theta_j)\Lambda^p(\theta_j+2\eta+i\pi),\quad j=1,\ldots,N,\\
&&\lim_{u\rightarrow \pm\infty}\Lambda^p(u)=(\frac{1}{2})^Ne^{\mp 3N\eta}[1+2\cosh(2M\eta)]+\cdots,\label{tsp3}
\eea
From these relations and the definition of $t^p(u)$, the eigenvalue $\Lambda^p(u)$ is a polynomial of $u$ with degree $2N$ and can be parameterized by its zeroes as
\bea
\Lambda^p(u)=\Lambda^p_0\prod_{j=1}^{2N}\sinh(\frac{u}{2}-\frac{z_j}{2}-\frac{3\eta}{2}),\label{Lamz-p}
\eea
Similarly, putting the parameterization (\ref{Lamz-p}) into (\ref{tsp1})-(\ref{tsp3}), we can deduce the homogeneous BAEs of the periodic IK model. The corresponding energy spectrum can be expressed as
\bea
E^p=&&\hspace{-0.6truecm}-2\frac{\partial \,\ln \Lambda^p(u)}{\partial u}\Big|_{u=0,\{\theta_j=0\}}\no\\
=&&\hspace{-0.6truecm}\sum_{j=1}^{2N}\coth(\frac{z_j}{2}+\frac{3\eta}{2}).\label{Energy-expressing-p}
\eea

In the ground state, all of zeroes form a set of bulk pairs $\{\bar{z}_j\sim \tilde{z}_j-5i\eta$, $(\tilde{z}_j-\pi)+5i\eta|j=1,\cdots,N\}$ with real $\{\tilde{z}_j\}$. Therefore, the energy spectrum (\ref{Energy-expressing-p}) at the ground state can be rewritten as
\bea
E^p=&&\hspace{-0.6truecm}i\sum_{j=1}^{N}[\cot(\frac{\tilde{z}_j}{2}-i\eta)+\tan(\frac{\tilde{z}_j}{2}-4i\eta)]\no\\
=&&\hspace{-0.6truecm}N\sum_{k=-\infty}^{\infty}[\tilde{a}_8(k)e^{-i\pi k}-\tilde{a}_2(k)]\tilde{\rho}^p(k).\label{Energy-expressing-pg}
\eea
where $\tilde{a}_n(k)=e^{-\eta|nk|}$, and $\tilde{\rho}^p(k)$ is the Fourier transformation of the continuum density $\rho^p(\tilde{z})$ for the $\tilde{z}$-zeroes. The function $\tilde{\rho}^p(k)$ can be solved by using the similar procedure to that applied in Section 5
\bea
\tilde{\rho}^p(k)=\frac{\tilde{b}_2(k)+(-1)^k\tilde{b}_4(k)}{1+(-1)^k\tilde{b}_6(k)},\label{rho-p}
\eea
where $\tilde{b}_n(k)=sign(k)ie^{-\eta|nk|}$. Finally, substituting (\ref{rho-p}) into (\ref{Energy-expressing-pg}), we have
\bea
E^p=2N\sum_{k=1}^{\infty}[(-1)^k e^{-\eta|6k|}-1]\frac{e^{-\eta|4k|}+(-1)^k e^{-\eta|6k|} }{1+(-1)^ke^{-\eta|6k|}}.
\eea
This completes the proof of the ground state energy (\ref{Ep}) for the periodic IK model.

\end{document}